\documentclass[aoas,preprint]{imsart}

\RequirePackage[OT1]{fontenc}
\RequirePackage{amsthm,amsmath}
\RequirePackage[numbers]{natbib}
\RequirePackage[colorlinks,citecolor=blue,urlcolor=blue]{hyperref}

\usepackage{graphicx}
\usepackage{tabularx}
\usepackage{url}
\arxiv{arXiv:2527566}

\startlocaldefs
\numberwithin{equation}{section}
\theoremstyle{plain}

\endlocaldefs

\begin{document}

\begin{frontmatter}
\title{Does the winning team \\always covers the point spread?\\ A study in professional basketball since 1990}
\runtitle{A study in professional basketball since 1990}

\begin{aug}
\author{\fnms{Luis A.} \snm{Mateos}}


\affiliation{MIT}



\end{aug}

\begin{abstract}
In sports betting it is easier to predict the winner of a game match than the team that covers the bet. Since, a winner team might not cover a bet.

This study focuses on the relation of the variable $win$ to the betting variable $cover$ the point spread. 
The study is performed with data from professional basketball (betting lines and scores) and tries to answer the question: 
Does the winning team always covers the point spread?. 
In order to answer this question, a regression analysis is performed taking into account the most and less winning teams, together with their betting variables since the 1990-1991 NBA season. The regression results are inserted in the SPXS expert system revealing an indirect factor analysis that correlates betting variables with teams winning percentages. 
\end{abstract}



\end{frontmatter}



Sports betting generally provides a predictable long-term advantage to the "house" or casino, while offering to the player a large short-term payout possibility. 
The player's disadvantage is a result of the casino not paying winning wagers according to the game's "true odds". In sports betting it is common that the bettor needs to win at least 52.4\% of the bets to be on a 50-50 fair game with the casino.

\begin{figure*}[t]
	\centering
	\includegraphics[width=0.99\textwidth]{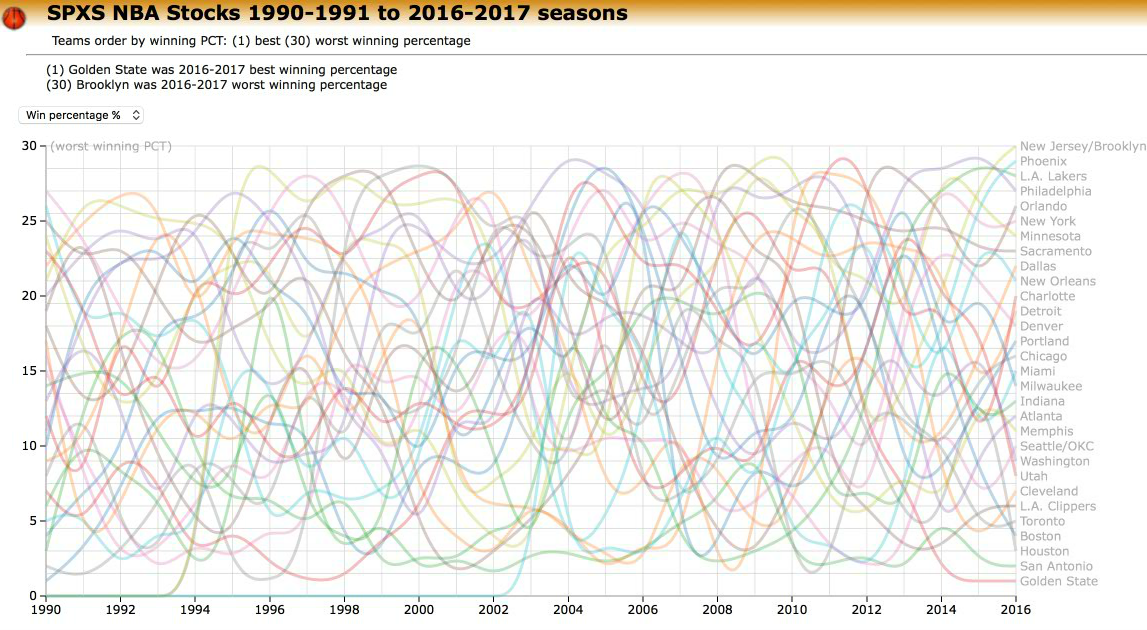}
	\caption{NBA teams sorted by winning averages (1990-1991 to 2016-2017 season). }
	\label{stocks}
\end{figure*}

Instead of trying to understand randomness in sports\cite{lopez} or apply competitive balance\cite{eco}\cite{angel} or performe cross-sport comparison\cite{david}. The presented paper goes back to the  fundamental analysis between the betting variables by applying an indirect factor analysis. Which is a statistical method used to describe variability among observed, correlated variables in terms of a potentially lower number of unobserved and generalized variables called indirect factors \cite{factora}. The study presents a novel and simple methodology to express the betting variables as a function of the team's winning percentage position, see Figure \ref{stocks}.

The paper is divided as follows: First, the model, data and expert system are introduced, in Section \ref{sec:tres} the regression results from the NBA seasons since 1990 are analyzed and in Section \ref{sec:cinco}, a "Player Edge" algorithm and strategy to overcome the house edge of 2.54\% is described.

\subsection*{Model}
The key assumption of the model is that there exist a relation of the teams with higher winning percentage to the teams that $cover$ the bet. And that there is a relation of the teams that lose the most (with the lowest winning average) to the teams that $no$ $cover$ the bet.

In this context, $cover$ the bet means that the team satisfies the betting line points or point spread and wins the bet. On the other hand, $no$ $cover$ the bet, means that the team does not satisfies the betting line points and loses the bet. 
First, let's define the betting variables:
\begin{itemize}
	\item $W/L$: Win or Lose a game match
	\begin{itemize}
		\item $win$: Win a game match
		\item $lose$: Lose a game match	
	\end{itemize}
	
	\item $C/N$: Cover or No cover the betting line points
	\begin{itemize}
		\item $cover$: Win a bet by covering the betting line points
		\item $no$ $cover$: Lose a bet by not covering the betting line points
		\item $push$: No action, no win neither lose
	\end{itemize}
	
	\item $O/U$: Over or Under the total points line
	\begin{itemize}
		\item $over$: Over the total points line 
		\item $under$: Under the total points line 
		\item $push$: No action, no over neither under
	\end{itemize}
\end{itemize}

The relation of winning a game match to $cover$ the bet is essential for answering the question: does the winning teams always cover the point spread?. This is because, it is easier to pick which team will win a game match instead of which team will cover the bet. The reason, is that $cover$ the point spread can be applied to a team that loses the game match.

In this sense, sports magazines and sportsbooks predict and make available their  ranking of teams before the NBA season starts. Hence, if the proposed relation exist between the variables $win$ and $cover$, this factor can help the bettor to counteract the house edge. 


\subsection*{Data}
The data for the presented study consists of all the betting lines and NBA game results since the 1990-1991 season to the 2013-2014. 
These records were collected from local sportsbooks in different countries, Mexico, US, UK and Austria. 
The data comprises hundreds of games divided in 24 seasons.

\subsection*{SPXS Expert System}
The SPXS \cite{spxs} - Sports Picks Expert System \cite{spxswww} is used for implementing a combinatorial regression to identify underlying correlations between the betting variables in the dataset. 


\begin{figure*}[t]
	\centering
	\includegraphics[width=0.99\textwidth]{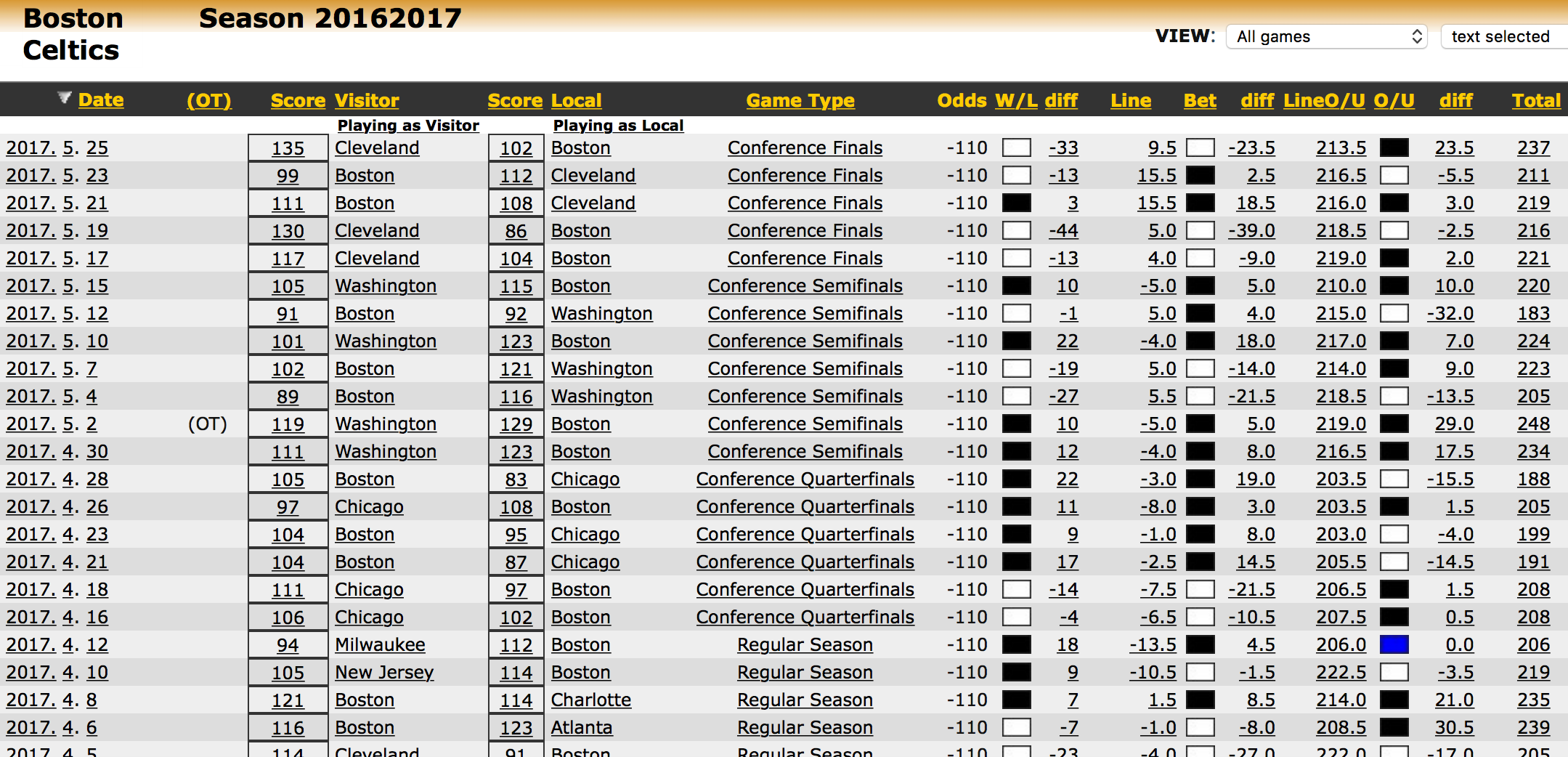}
	\caption{DNA visualization of betting variables and their differences or error (Boston in the 2016-2017 NBA season).}
	\label{ffig2}
\end{figure*}

The expert system analyzes the three main variables that results from a game match: $W/L$ ($win/lose$) the game match; $C/N$ ($cover/no$ $cover$) the betting line (point spread); $O/U$ ($over/under$) total points line (TPL)  and their differeces or error   to the casino's betting lines, see Figure \ref{ffig2}. These deltas are: $\Delta w$ is the difference of points from the final score in a game match between two teams, $\Delta c$ is the error in points with respect to the favorite team to win the game match (ATS), and $\Delta o$ is the error in points with respet to the TPL.


\section{Regression}
\label{sec:tres}

The teams winning percentages in the analyzed NBA seasons (1990-1991 to 2013-2014) ranged from $\approx25\%$  to $\approx70\%$. Therefore, the value range among the spectrum of teams is well defined and easier to differentiate between groups of teams: Teams with low winning percentage can be defined as teams $W_{pct}<0.3450$; Team with high winning percentage can be defined as teams $W_{pct}>0.6400$; Teams with mediocre winning percentage are the teams in between the high and low groups.

For example, in the NBA season 2010-2011, see Table \ref{w20112012}. It is possible to identify the six teams with low winning percentages $\bar W_{pct} \approx 26\%$ (team positions 1-6), the seven teams with high winning percentages $\bar W_{pct} \approx 71\%$ (team positions 24-30). And the teams with mediocre winning percentage (team positions 7-23). 

The regression analysis includes the NBA seasons from 1990-1991 to 2013-2014 subdivided  in 3 subgroups, defined by the number of teams per season. The first subgroup includes the seasons 1990-1991 to 1994-1995 when the NBA has only 27 teams, the second subgroup includes the seasons 1995-1996 to 2003-2004 when the NBA has 29 teams, and the last subgroup includes the seasons from 2004-2005 to the 2013-2014 when the NBA started to have 30 teams.

\textbf{Note} that this regression study only comprises the games in regular season from 1990-1991 to 2013-2014 as the "training set" and the NBA seasons 2014-2015 to 2015-2016 are the "test set".


\subsection*{Methodology}

The key and novelty of the presented methodology is to analyze the data by taking into account the \textbf{position of the teams from their winning percentages}  to find a factor analysis that correlates the team's position to the betting variables. 

Let's define the variables to analyze:
\begin{itemize}
	\item $T_{Hw_{pct}}$: Teams with high winning percentages
	\item $T_{Lw_{pct}}$:  Teams with low winning percentages
	\item $t_{Hw_{pct}}$: Team with highest winning percentages
	\item $t_{Lw_{pct}}$:   Team with lowest winning percentages
	\item $T_{Hc_{pct}}, T_{Lc_{pct}}, t_{Hc_{pct}}, t_{Lc_{pct}}$: for covering the point spread percentages respectively
	\item $T_{Ho_{pct}}, T_{Lo_{pct}}, t_{Ho_{pct}}, t_{Lo_{pct}}$: for over the total points line percentages respectively
\end{itemize}

\begin{table}[]
	\centering
	\caption{Teams sorted by winning percentages in the NBA 2010-2011 season (Regular season with 82 games played per team).}
	\label{w20112012}
	\begin{tabular}{|l|l|l|l|l|}
		\hline
		Team position & Team      & W  & L  & Percentage \\ \hline
		1& Minnesota & 17 & 65 & 0.2073     \\ \hline
		2& Cleveland          &  19  &  63  &  0.2317          \\ \hline
		3& Toronto          &  22  &  60  &   0.2683         \\ \hline
		4&  Washington         &  23  &  59  &   0.2805         \\ \hline
		5&  Sacramento         &  24  & 58   &  0.2927          \\ \hline
		6&  New Jersey         &  24  &  58  &    0.2927        \\ \hline \hline
		
		7&  Detroit         & 30   &  52  &   0.3659         \\ \hline
		8&  LA Clippers         &  32  &  50  &    0.3902        \\ \hline
		9&  Charlotte         &  34  &  48  &  0.4146          \\ \hline
		10& Milwaukee          &  35  &  37  &   0.4268         \\ \hline
		11& Golden State & 36 & 46 & 0.4390     \\ \hline
		12&   Indiana        &  37  & 45   &   0.4512         \\ \hline
		13&     Utah      &  39  &  43  &    0.4756        \\ \hline
		14&   Phoenix        & 40   &  42  &   0.4878         \\ \hline
		15&   Philadelphia        &  41  &  41  &   0.5000         \\ \hline
		16&   New York        & 42   & 40   &   0.5122         \\ \hline
		17&    Houston       &  43  &  39  &    0.5244        \\ \hline
		18&    Atlanta       &  44  & 38   &   0.5366         \\ \hline
		19&    Vancouver       &  46  &  36  &   0.5610         \\ \hline
		20&    New Orleans       & 46   &  36  &   0.5610         \\ \hline
		21 & Portland & 48 & 34 & 0.5854     \\ \hline
		22&  Denver         &  50  &  32  &  0.6098          \\ \hline
		23&  Orlando         &  52  & 30   &   0.6341         \\ \hline \hline
		
		24&   Oklahoma City        &  55  & 27   &  0.6707          \\ \hline 
		25&  Boston         & 56   & 26   &   0.6829         \\ \hline
		26&  Dallas         & 57   &  25  &   0.6951         \\ \hline
		27&   LA Lakers        & 57   &  25  &  0.6951          \\ \hline
		28&   Miami        &  58  &  24  &   0.7073         \\ \hline
		29&   San Antonio        & 61   &  21  &   0.7439         \\ \hline
		30&   Chicago        & 62   &  20  &   0.7561         \\ \hline
	\end{tabular}
\end{table}

And state the key questions: 

\begin{itemize}
	\item How much does the variable $win$ varies? 
	
	\item How much does the variable $cover$ the point spread varies? 
	
	\item How much does the variable $over$ the TPL varies?
	
	\item How much does the variable $cover$ the point spread varies with respect to the  teams $winning$ percentage position?
	
	\item How much does the variable $over$ the TPL varies with respect to the  teams $winning$ percentage position?
	
	
\end{itemize}

The first analytic factor to understand is how much the teams winning percentages vary in the data set. From each NBA season the teams are sorted from their winning percentages as shown in Table \ref{w20112012} and average together for each NBA seasons subgroup. 

\subsection{How much does the variable $win$ varies? }
\begin{figure}[t]
	\centering
	\includegraphics[width=0.9\textwidth]{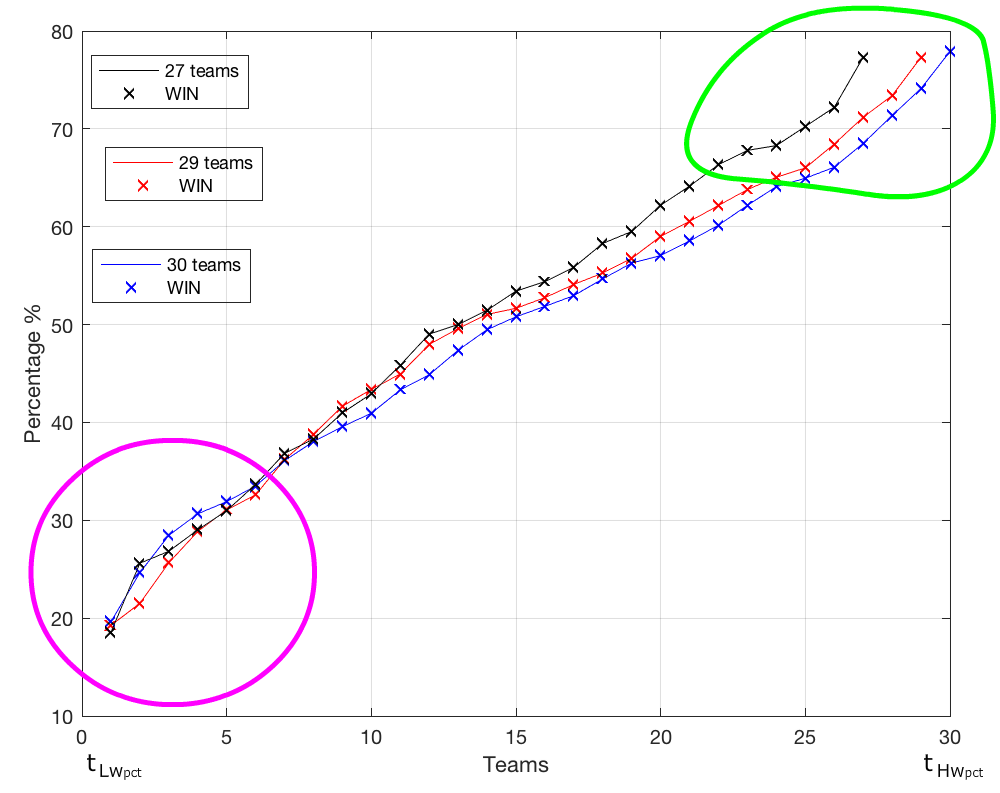}
	\caption{Betting variable $win$ sorted by teams winning percentages (1990-1991 to 1994-1995 seasons (27 teams);  1995-1996 to 2003-2004 seasons (29 teams); 2004-2005 to 2013-2014 seasons (30 teams)).}
	\label{w27w}
\end{figure}

Figure \ref{w27w}, shows the plot from the $win$ variable sorted by wining percentages, revealing that the  teams with high winning percentages have in average $T_{Hw_{pct}}\approx70\%$ while the  teams with low winning percentages have around $T_{Lw_{pct}}\approx25\%$. 

Also, Figure \ref{w27w} shows a similar plot shape in the three subgroups. The teams with highest winning percentages are nearly $\approx80\%$ and the teams with lowest winning percentages are below $\approx20\%$.

In this sense, the $win$ variable is easier to differenciate if taking into account not only the teams with the best $t_{Hw_{pct}}$ and the worst $t_{Lw_{pct}}$ winning percentage. But also, if taking a couple of groups (i.e., the six most winning teams and their counterpart, the six teams with lower winning percentage). 

\subsection{How much does the variable $cover$ the "point spread" varies? }
Figure \ref{c27c}, shows the plot of the variable $cover$ sorted by the teams winning percentage position, so the representing teams in the position 1 are the teams $t_{Lw_{pct}}$ with the  lowest winning percentage and the team in position 27, 29 and 30 represents the teams $t_{Hw_{pct}}$ with the highest winning percentages in each subgroup of NBA seasons. 

It is possible to notice that the pattern of Figure \ref{c27c} follows the shape of the winning percentages plot from Figure \ref{w27w}. Eventhough, the plot from the $cover$ variable shows high frequency changes. However, if taking into account the outliers shapes, the teams with high covering percentage $T_{Hc_{pct}}\approx T_{Hw_{pct}}$ approximate the teams with high winning percentages and the  teams with low covering percentage $T_{Lc_{pct}}\approx T_{Lw_{pct}}$ approximate the teams with low winning percentages. 

\subsection{How much does the variable $over$ the total points line (TPL) varies? }
The variable $over$ shows an inverted behaviour from the $cover$ and $win$ variables. Figure \ref{c27c}, shows the plot of the variable $over$ sorted by the teams with their winning percentages. Similar to the previous plot, the team in the position 1 is the one with the lowest winning percentages over the seasons $t_{Lw_{pct}}$ and $t_{Hw_{pct}}$ is the highest. 

Even if the plot reveals a pattern, most of the points fall within the range of the house edge of $50\%\pm2.54\%$ (highlighted in gray area in Figure  \ref{c27c}). 



\subsection{How much does the variable $cover$ the "point spread" varies with respect to $win$? }

The results reveal a factor analysis between the $win$ and the $cover$ variables. 
The $cover$ variable tends to follows on the positive outliers from the $T_{Hw_{pct}}$ teams with the best winning percentages and tends to follow on the negative outlier from the $T_{Lw_{pct}}$ teams with the worst winning percentages. While, the teams with mediocre winning percentages remain inside the house edge threshold of $\pm$2.4\%.

\begin{figure}[h]
	\centering
	\includegraphics[width=0.49\textwidth]{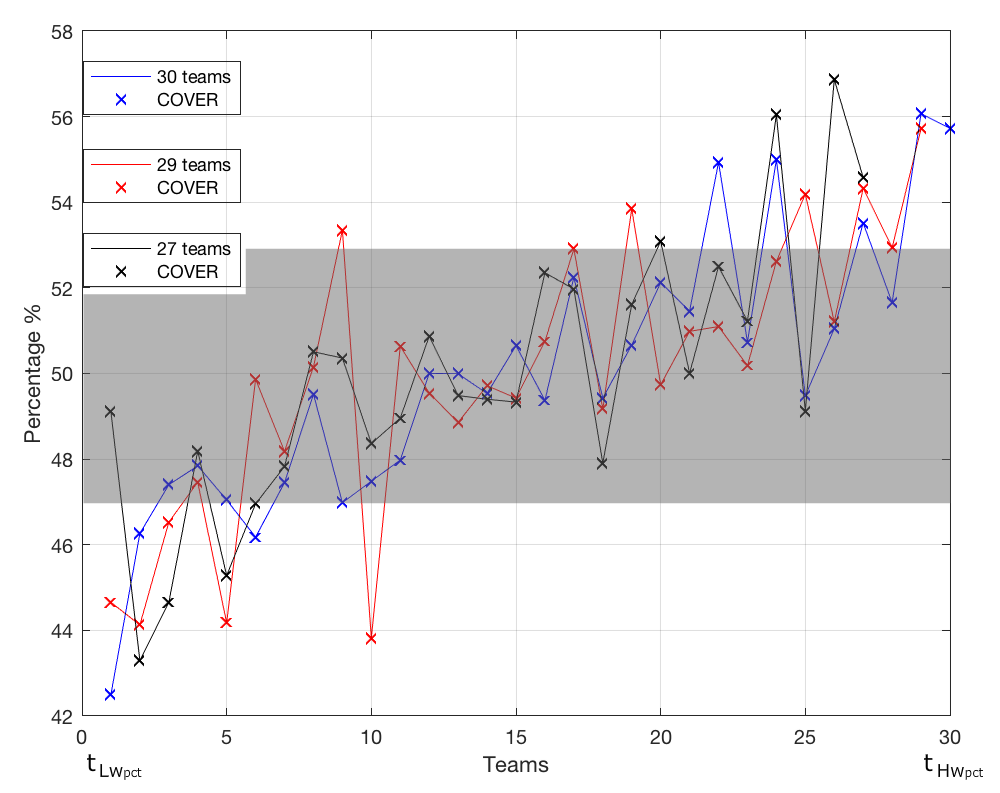}
	\includegraphics[width=0.49\textwidth]{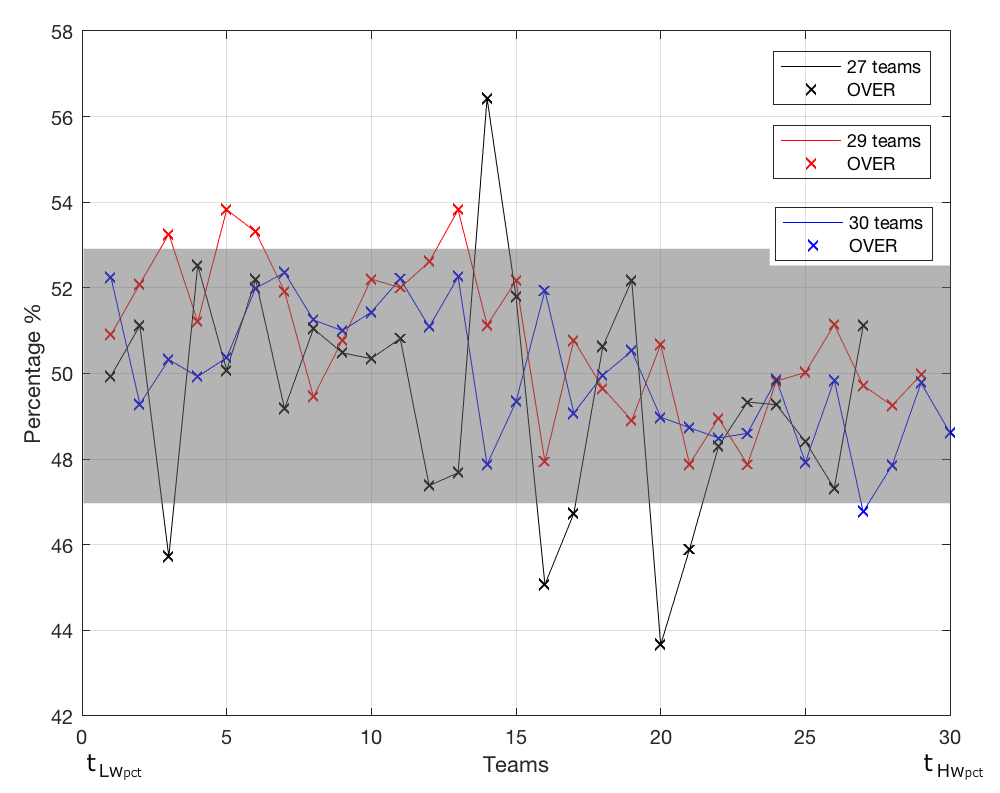}
	\caption{Betting variables $cover$ the point spread and $over$ the total points line. All sorted by teams winning averages (1990-1991 to 1994-1995 seasons (27 teams);  1995-1996 to 2003-2004 seasons (29 teams); 2004-2005 to 2013-2014 seasons (30 teams)). (Gray area is defined as the house edge of $\pm2.54\%$)}
	\label{c27c}
\end{figure}

However, the tendency $win \leftrightarrow  cover$ is not straightfoward. 
Figure \ref{c27c}, shows the results for the seasons 1990-1991 to 1994-1995 (27 teams) revealing a positive outliers for the two most winning teams. But, for the $t_{Lw_{pct}}$ team with the lowest winning percentage there is no outlier as its $cover$ variable falls inside the house edge, around $\approx$49$\%$.

\section{Player Edge - Algorithm and Strategy}
\label{sec:cinco}

From the previous section, it is easy to notice that selecting who is going to win a game match is easier that selecting who is going to $cover$ the point spread. In the same way, it is easier to select who is going to lose the match that select who is going to $no$ $cover$ the point spread.

In order to use these results in a consistent way, a running average algorithm is integrated in the formulation. The objective of the algorithm is to smooth the outlier so they compensate each other while still obtaining a continuos outlier shape in the two ends of the plot. This technique is used in the stock market for defining stocks trends \cite{Knuth:1997:ACP:270146}.





\subsection{Algorithm}

The process is as follows: the team's sorted winning percentage position is taken to obtain the covering percentage of the position in consideration. 

The algorithm is performed on the two outliers, meaning that one is for the $T_{Hw_{pct}}$ teams with high winning percentages starting on the $t_{Hw_{pct}}$ team with the highest winning average, moving to the second highest and so on,  until the running average fails to overcome the house edge of 2.54\%, see Equation \ref{equ1}, where $\bar{H}_{Cpct}$ is the value of the running average for the $cover$ variable with respect to the team's winning percentage position, $t_{c_{pct}}$ is the team's covering percentage. 


\begin{equation}
\bar{H}_{Cpct} = \frac{1}{T_{Hw_{pct}}}\sum\limits_{i=0}^{T_{Hw_{pct}}-1} t_{c_{pct}}(t_{Hw_{pct}}) - i
\label{equ1}
\end{equation}

In the same way, the second instance of the algorithm is applied to the $T_{Lw_{pct}}$ teams with low winning percentages, starting on the $t_{lw_{pct}}$ team with the lowest winning percentage and moving to the second lowest and so on until the running average fails to overcome the house edge, see Equation \ref{equ2},  $\bar{H}_{Npct}$ is the value of the running average for the $no$ $cover$ variable with respect to the team's position winning percentage, $t_{n_{pct}}$ is the team's no covering percentage. 


\begin{equation}
\bar{H}_{Npct} = \frac{1}{T_{Lw_{pct}}}\sum\limits_{i=0}^{T_{Lw_{pct}}-1} t_{n_{pct}}(t_{Lw_{pct}}) - i
\label{equ2}
\end{equation}


\subsection*{Cover the point spread variable}
\begin{figure*}[thpb]
	\centering
	\includegraphics[width=0.92\textwidth]{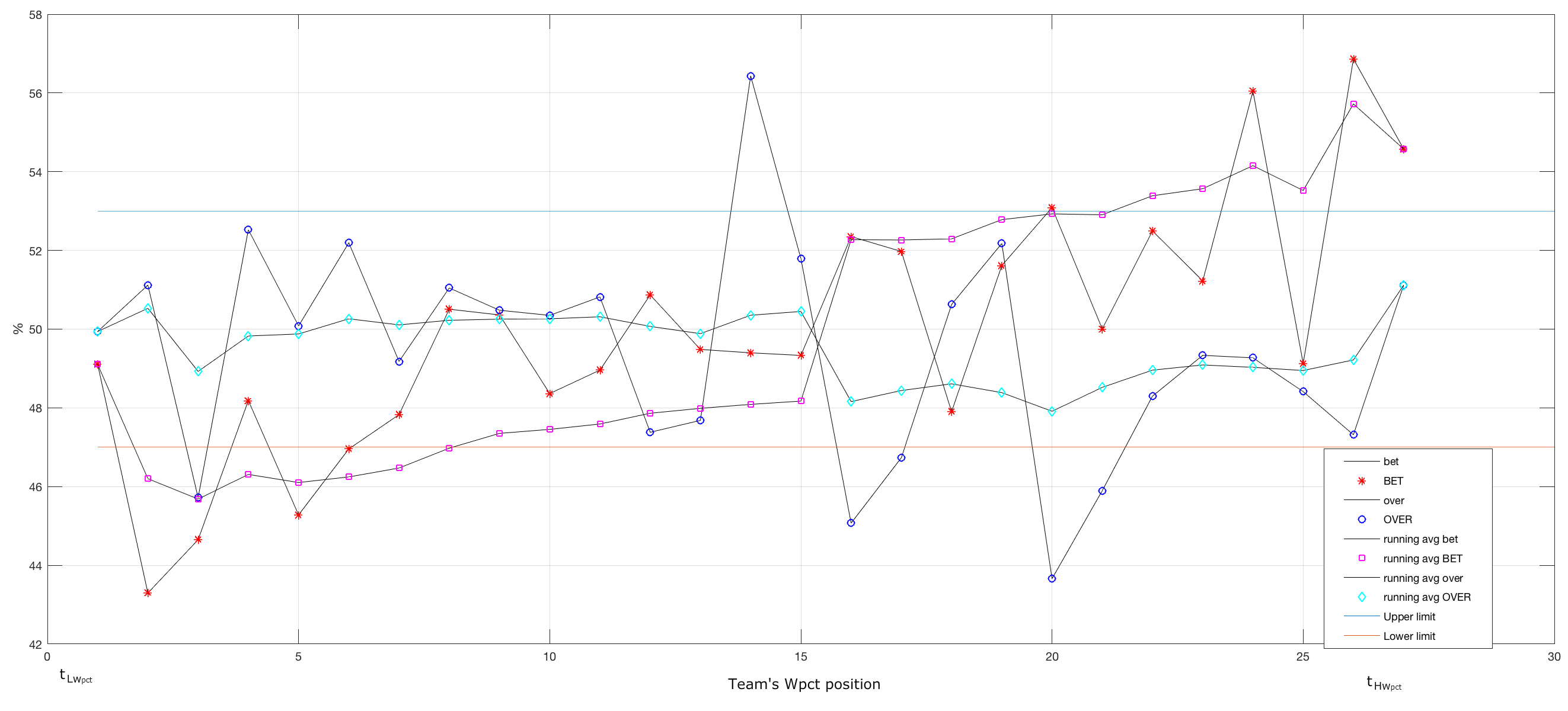}
	\caption{Running avg. algorithm in NBA Seasons 1990-1991 to 1994-1995 (27 teams). }
	\label{fig27teams}
\end{figure*}

The results from applying the running average algorithm to the first subgroup, which includes the 1990-1991 to 1994-1995 NBA seasons (27 teams), are  shown in Figure \ref{fig27teams}. In this subgroup, if selecting the six best teams with winning percentages will lead to overcome the house edge on the positive end of  +2.54\%. And, if selecting the worst seven winning percentages teams will overcome the house edge on the negative end of -2.54\%. 

\begin{figure*}[thpb]
	\centering
	\includegraphics[width=0.92\textwidth]{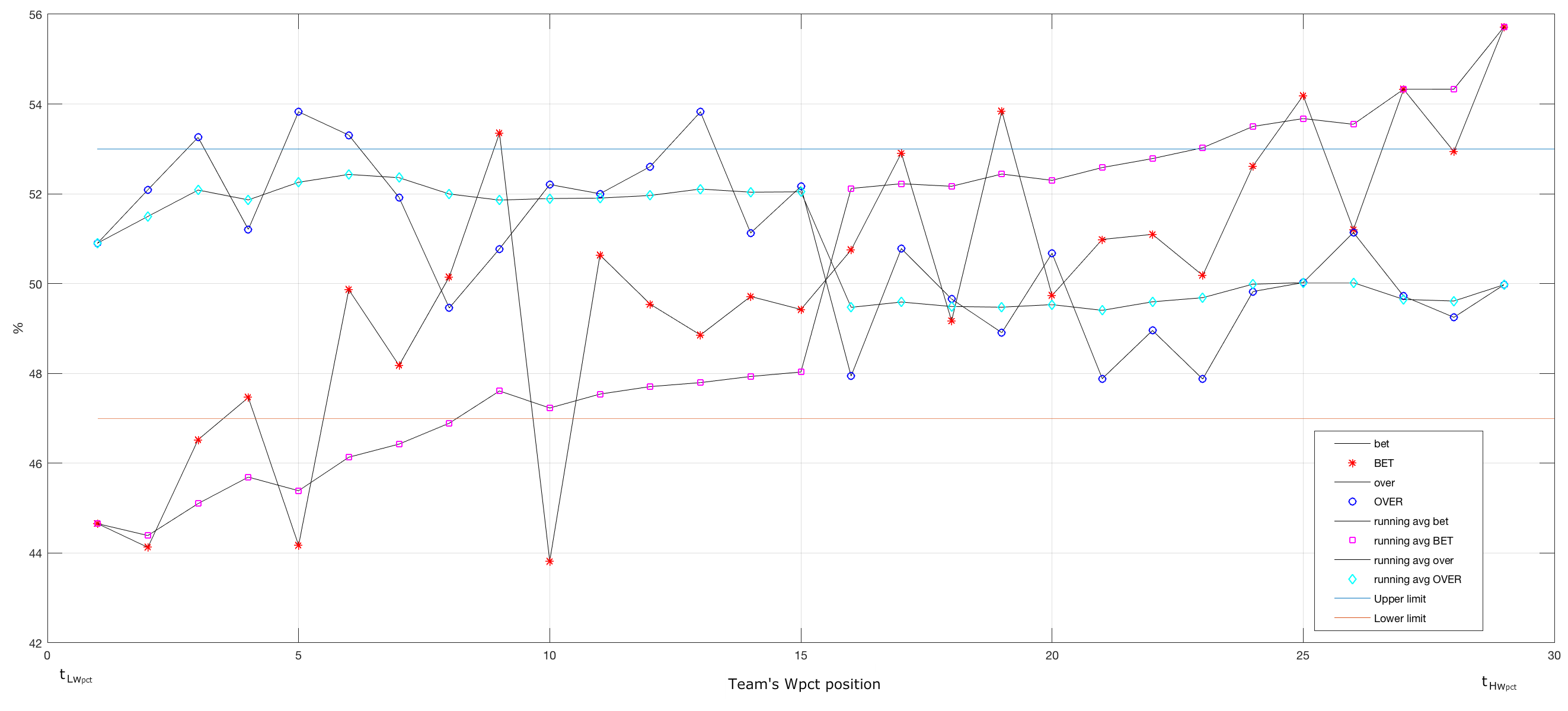}
	\caption{Running avg. algorithm in NBA Seasons 1995-1996 to 2003-2004 (29 teams). }
	\label{fig29teams}
\end{figure*}

The results from applying the moving average algorithm to the second soubgroup, which includes the 1995-1996 to 2003-2004 NBA seasons with 29 teams, are shown in Figure \ref{fig29teams}. The results reveal that the seven best teams with winning percentages surpassed the house edge +2.54\%. And the eight teams with the lowest winning percentages also surpassed the -2.54\% threshold.

\begin{figure*}[thpb]
	\centering
	\includegraphics[width=0.92\textwidth]{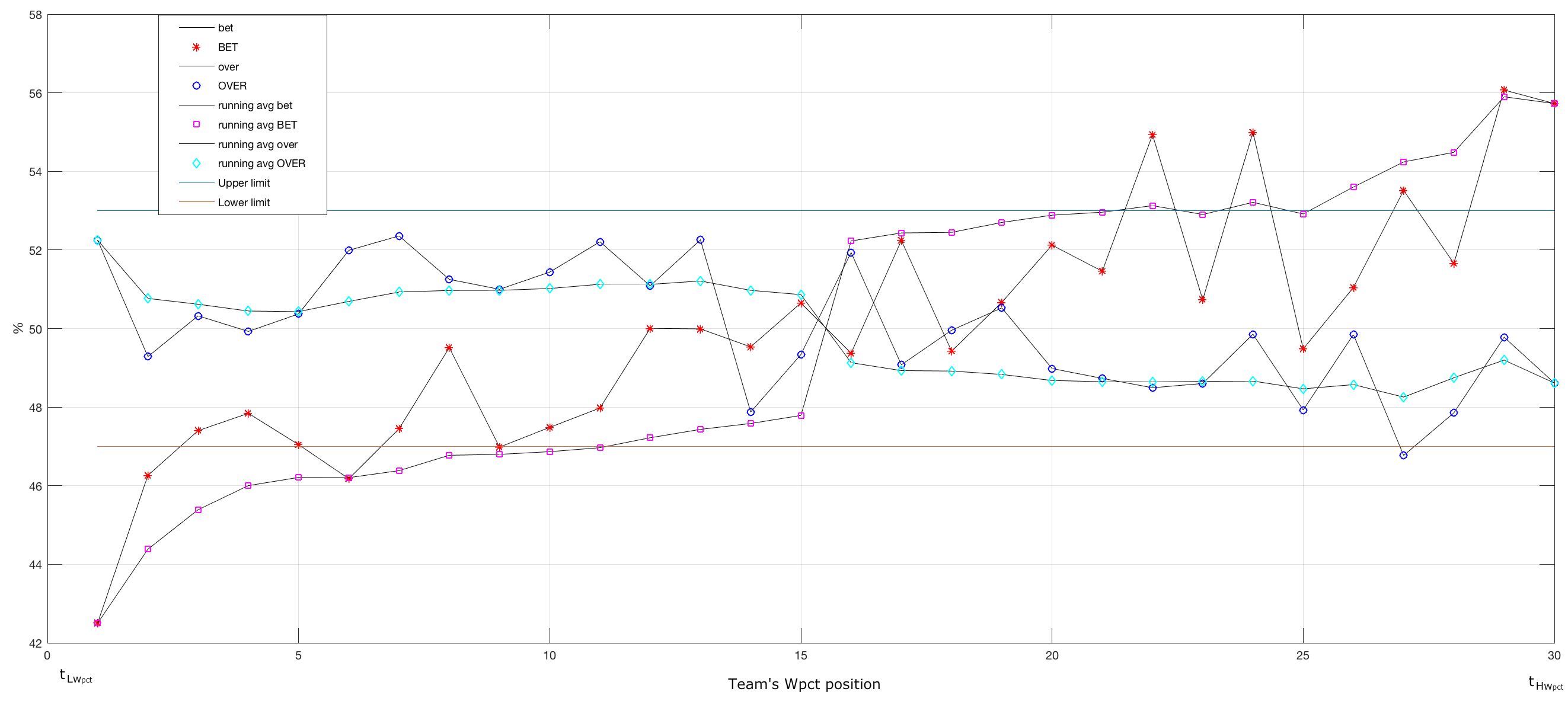}
	\caption{Running avg. algorithm in NBA Seasons 2004-2005 to 2013-2014 (30 teams). }
	\label{fig30teams}
\end{figure*}

In the current $era$ of professional basketball with 30 teams, comprising the 2004-2005 to 2013-2014 NBA seasons, the results from applying the moving average algorithm show that the five teams with highest winning percentages and the ten teams with worst winning percentages surpassed the house edge $\pm2.54\%$, see Figure \ref{fig30teams}.

\subsection*{Over the total points line (TPL) variable}
Applying the running average algorithm to the $over$ the TPL variable results in a plot that remains in the house edge domain, see Figure \ref{fig27teams}, \ref{fig29teams} and \ref{fig30teams}. The only notable pattern is a tendency of the $T_{Hw_{pct}}$  teams with high winning percentages to have an $under$ $\approx$1.3\% of the times more likelly than an $over$. And a tendency of the $T_{Lw_{pct}}$ teams with low winning percentages to have an $over$ $\approx$1.2\% of the times more likelly than an $under$.

\begin{figure*}[t]
	\centering
	\includegraphics[width=0.49\textwidth]{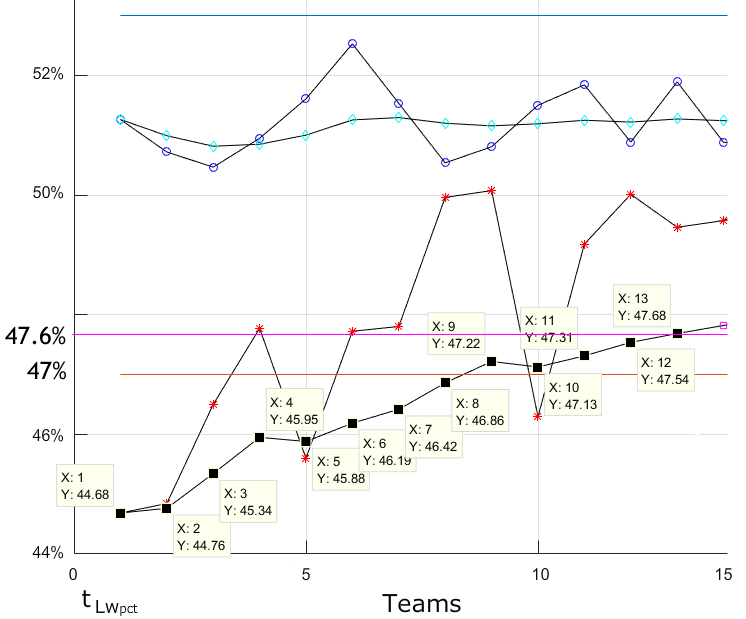}
	\includegraphics[width=0.49\textwidth]{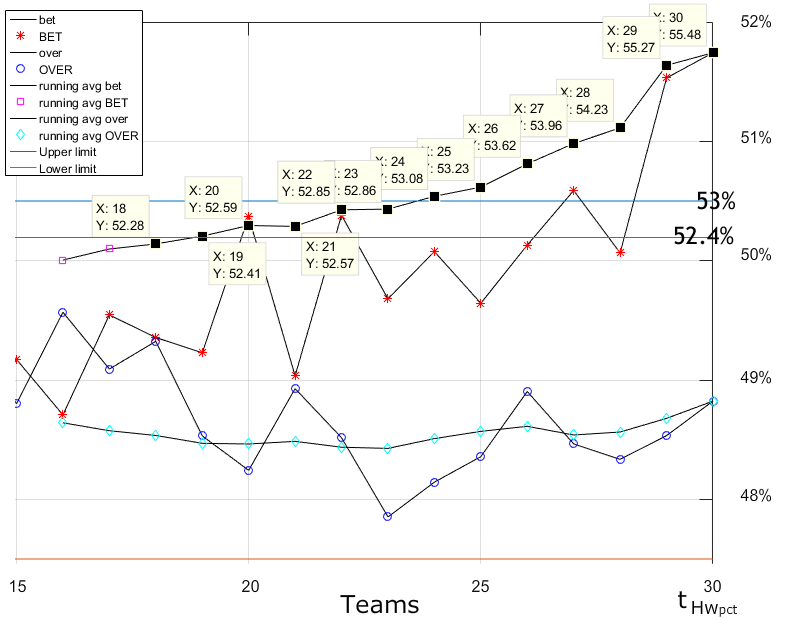}
	\caption{ Running average during the NBA seasons 1990-1991 to 2013-2014 from the teams with lowest winning percentages $t_{Lw_{pct}}$ (left) and from the teams with highest winning percentages $t_{Hw_{pct}}$ (right) revealing the set of teams that surpassed the house edge of $\pm2.54\%$.}
	\label{fighighteams}
\end{figure*}

\subsection{Strategy}
From the regression study in Section \ref{sec:tres} together with the proposed algorithm, a couple of correlations emerged between the variables $W/L$  $\leftrightarrow$ $C/N$ and $W/L$  $\sim \leftrightarrow$ $O/U$. The correlation coefficient assume values in the range from -1 to +1, where +1 indicates the strongest possible agreement and -1 the strongest possible disagreement \cite{corr}. 
This correlation is calculated as the proportion of the extreme outliers from the variables in consideration:

\begin{itemize}
	\item $W/L$  correlation coefficient to $C/N$ is $\approx$ 0.2
	\item $W/L$  correlation coefficient to $O/U$ is $\approx$ -0.04
\end{itemize}

\begin{table}[t]
	\centering
	\caption{Less winning teams in NBA seasons 1990-1991 to 2013-2014}
	\label{lesswinning}
	\begin{tabular*}{0.9\textwidth}{@{\extracolsep{\fill} }c c c r }
		\hline
		No cover the "point spread" & $\bar{H}_{Npct}$ & Profit &$\sum$ \\  \hline
		The most losing team & 44.68\%    & 2.92\%  &~\\ \hline
		The $2^{nd}$ most losing teams & 44.76\%    & 2.84\%&~  \\ \hline
		The $3^{rd}$ most losing teams & 45.34\%    & 2.26\% &~ \\ \hline
		The $4^{th}$ most losing teams & 45.95\%    & 1.65\% &~ \\ \hline		
		The $5^{th}$ most losing teams & 45.88\%    & 1.72\% &~ \\ \hline
		The $6^{th}$ most losing teams & 46.19\%    & 1.41\% &~ \\ \hline
		The $7^{th}$ most losing teams & 46.42\%    & 1.18\% &~ \\ \hline
		The $8^{th}$ most losing teams & 46.86\%    & 0.74\% &~ \\ \hline
		The $9^{th}$ most losing teams & 47.22\%    & 0.38\% &~ \\ \hline
		The $10^{th}$ most losing teams & 47.13\%    & 0.47\% &~ \\ \hline
		The $11^{th}$ most losing teams & 47.31\%    & 0.29\% &~ \\ \hline
		The $12^{th}$ most losing teams & 47.54\%    & 0.06\% &$\sum = 15.92\%$ \\ \hline
	\end{tabular*}
\end{table}

The strategy consist in taking groups of teams from the two outliers, one $T_{Hw_{pct}}$: Teams with high winning percentages and the second for the $T_{Lw_{pct}}$:  Teams with low winning percentages that surpassed the house edge. 

Figure \ref{fighighteams} shows the complete study for the "training set", the 24 NBA season (1990-1991 to 2013-2014) and defines visually the groups of teams, $T_{Hw_{pct}}$: Teams with high winning percentages as the best twelve teams with the highest winning percentage to cover the point spread.
and $T_{Lw_{pct}}$:  Teams with low winning percentages as the best twelve teams with the lowest winning percentage to fail to cover the point spread. The groups are defined from the extreme values $t_{Hw_{pct}}$ and $t_{Lw_{pct}}$ untill the running average algorithm falls in the house edge of $50\pm2.54\%$.


\begin{table}[t]
	\centering
	\caption{Most winning teams in NBA seasons 1990-1991 to 2013-2014}
	\label{mostwinning}
	\begin{tabular*}{0.9\textwidth}{@{\extracolsep{\fill} } l l l l }
		\hline
		Cover the "point spread" & $\bar{H}_{Cpct}$ &  Profit & $\sum$\\  \hline
		The most winning team & 55.48\%    & 3.08\% &~ \\ \hline
		The $2^{nd}$ most winning teams & 55.27\%    & 2.87\% &~  \\ \hline
		The $3^{rd}$ most winning teams & 54.23\%    & 1.83\%  &~ \\ \hline
		The $4^{th}$ most winning teams & 53.96\%    & 1.56\% &~  \\ \hline		
		The $5^{th}$ most winning teams & 53.62\%    & 1.22\% &~  \\ \hline
		The $6^{th}$ most winning teams & 53.23\%    & 0.83\% &~  \\ \hline
		The $7^{th}$ most winning teams & 54.08\%    & 0.68\% &~  \\ \hline
		The $8^{th}$ most winning teams & 52.86\%    & 0.46\% &~  \\ \hline
		The $9^{th}$ most winning teams & 52.85\%    & 0.45\% &~  \\ \hline
		The $10^{th}$ most winning teams & 52.57\%    & 0.17\% &~  \\ \hline
		The $11^{th}$ most winning teams & 52.59\%    & 0.19\% &~  \\ \hline
		The $12^{th}$ most winning teams & 52.41\%    & 0.01\% & $\sum = 13.35\%$  \\ \hline
	\end{tabular*}
\end{table}

\begin{figure*}[t]
	\centering
	\includegraphics[width=0.99\textwidth]{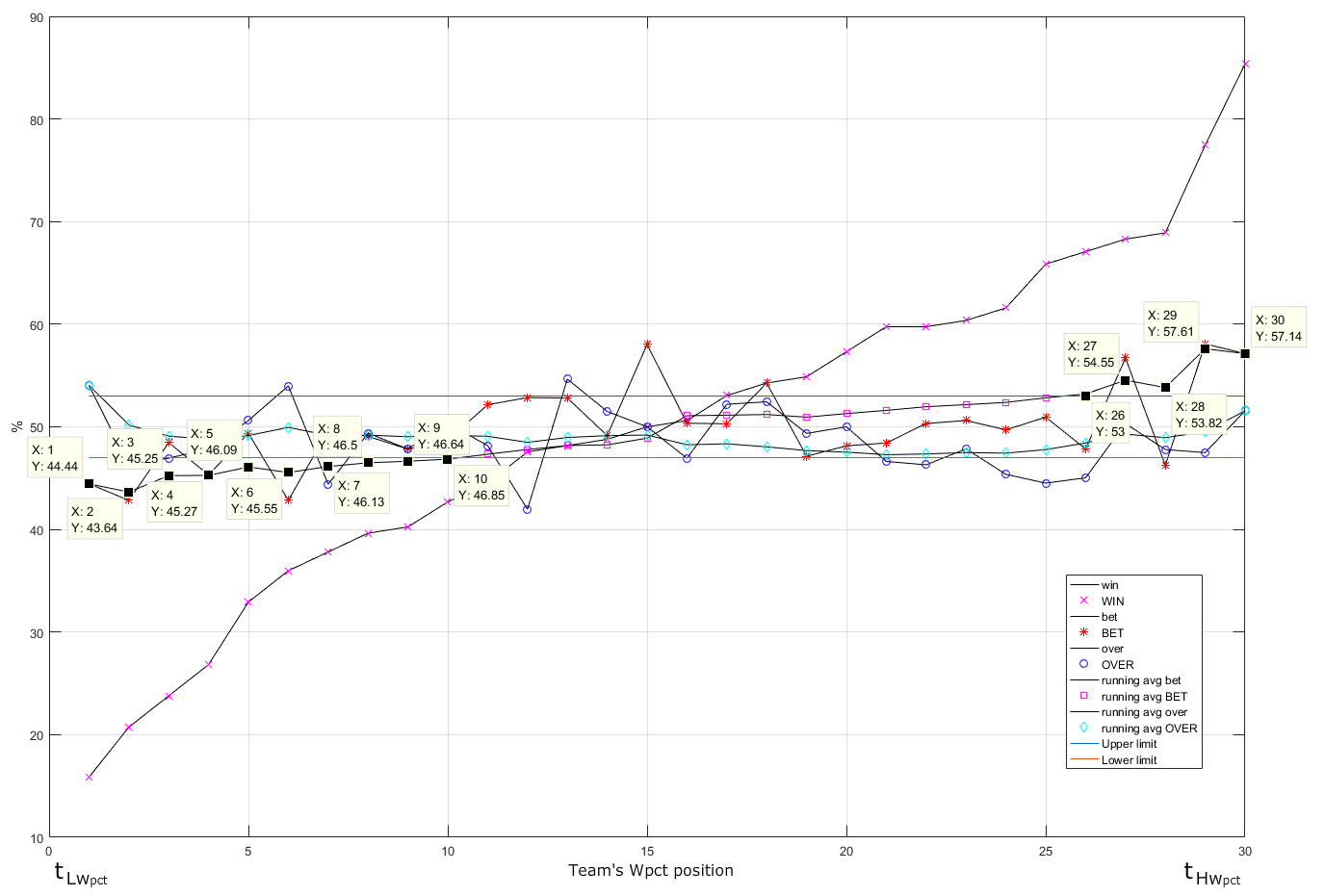}
	\caption{ Player edge algorithm highligthing team positions that surpassed the house edge for the NBA seasons 2014-2015 and 2015-2016 used as "test set".}
	\label{fighigh1415}
\end{figure*}

Table \ref{lesswinning}, shows the $T_{Lw_{pct}}$ as the twelve low winning percentages teams that surpassed the casino's edge on the negative end. 
In the same way, Table \ref{mostwinning}, shows the $T_{Hw_{pct}}$ as the twelve high winning percentages teams that surpassed the casino's edge. 
The summation of percentages reveals that it is a better option to select the $T_{Lw_{pct}}$ teams with low winning percentage to $no$ $cover$ the point spread, since their culmulative percentage of 15.92\% is higher than to the   $T_{Hw_{pct}}$ with 13.35\%.

From the "training set" results, let's define a threshold of $profit$ $>2\%$ to test the Player Edge strategy. 
In this sense, the three teams with the lowest winning percentages and the two teams with the highest are considered in the strategy, see Tables \ref{lesswinning} and \ref{mostwinning}.

In order to test the performance of the proposed Player Edge strategy, the NBA 2014-2015 and 2015-2016 seasons are used as "test set". Figure \ref{fighigh1415} shows the data from applying the Player Edge algorithm, resulting in profits for the selected range, see Tables \ref{lesswinningtest} and \ref{mostwinningtest}. Also, Figure \ref{fighigh1415} shows the Player Edge algorithm until it converges on the two ends. For the $T_{Lw_{pct}}$ teams with lowest winning percentages the worst ten team's percentages surpassed the house edge, while for the $T_{Hw_{pct}}$ teams with higher winning percentages, only the five most winning team's percentages surpassed the house edge on the positive end. In \cite{pewww} a detailed information of the Player Edge algortihm applied to each NBA season can be found. 

\begin{table}[t]
	\centering
	\caption{Less winning teams in NBA seasons 2014-2015 to 2015-2016}
	\label{lesswinningtest}
	\begin{tabular*}{0.9\textwidth}{@{\extracolsep{\fill} }c c c r }
		\hline
		No cover the "point spread" & $\bar{H}_{Npct}$ & Profit &$\sum$ \\  \hline
		The most losing team & 44.44\%    & 3.16\%  &~\\ \hline
		The $2^{nd}$ most losing teams & 43.64\%    & 3.96\%&~  \\ \hline
		The $3^{rd}$ most losing teams & 45.25\%    & 2.35\% &$\sum = 9.47\%$ \\ \hline
	\end{tabular*}
\end{table}

\begin{table}[t]
	\centering
	\caption{Most winning teams in NBA seasons 2014-2015 to 2015-2016}
	\label{mostwinningtest}
	\begin{tabular*}{0.9\textwidth}{@{\extracolsep{\fill} } l l l l }
		\hline
		Cover the "point spread" & $\bar{H}_{Cpct}$ &  Profit & $\sum$\\  \hline
		The most winning team & 57.14\%    & 4.74\% &~ \\ \hline
		The $2^{nd}$ most winning teams & 57.61\%    & 5.21\% &$\sum = 9.95\%$  \\ \hline
	\end{tabular*}
\end{table}

\section{Conclusion and future work}
\label{sec:conclusion}


This paper presents a study in professional basketball NBA with the aim to answer the question: Does the winning team always covers the bet?. 

In order to answer the question, the SPXS - Sports Picks eXpert System is used for performing the regression analysis of all the NBA game matches in regular season, from the  1990-1991 to 2015-2016 season. 
This dataset is divided and analyzed in two subsets similar to the machine learning - supervised data mining algorithms: 

A "training set" covering the NBA seasons from 1990-1991 to 2013-2014 is used for revealing an indirect factor analysis within the betting variables: $cover$ the point spread and $over$  the total points line (TPL) from the team's winning percentage position. 

And a "test set" covering the NBA seasons 2014-2015 to 2015-2016 is used for corroborating the hypothetical indirect factor analysis between betting variables. 

Also, a $Player$ $Edge$ algorithm and strategy is described, showing a methodology for a possible long-term advantage to the player to surpass the house edge of 2.54\%

Moreover, the presented indirect factor analysis can be applied to stock market for defining stocks trends from indirect generalized variables 
and also for DNA analysis to relate patterns from the indirect  generalization of well defined variables.



\begin{thebibliography}{9}

\bibitem{Knuth:1997:ACP:270146}
\textsc{Knuth, Donald E.} (1997). \textit{The Art of Computer Programming, Volume 2 (3rd Ed.): Seminumerical Algorithms}, Addison-Wesley.  Boston, MA, USA

\bibitem{spxswww}
\textsc{Luis A. Mateos}  (2018-01-03). \textit{Expert System SPXS website , accessed: 2018-01-03}  
\textcolor{blue}{http://www.particlerobots.com/spxs/}

\bibitem{spxs}
\textsc{Luis A. Mateos} (2011).
\textit{SPXS Sports Picks eXpert System}.
ICAI 2011, Las Vegas.

\bibitem{pewww}
\textsc{Luis A. Mateos}  (2018-01-03). \textit{Player Edge applied to NBA 1990-1991 to 2015-2016, 2018-01-03}  
\textcolor{blue}{http://www.particlerobots.com/luismateos/spxs/appendix.html}

\bibitem{factora}
\textsc{Child, Denni} (2006).
The Essentials of Factor Analysis \textit{Contemporary Sociology ; 10.2307/2061984}

\bibitem{lopez}
\textsc{Lopez, Michael} (2017).
How often does the best team win? A unified approach to understanding randomness in North American sport \textit{ arXiv:1701.05976 [stat.AP]}

\bibitem{angel}
\textsc{Angel Barajas} (2014).
Reinventing the economics of sport \textit{Sport, Business and Management: An International Journal; 10.1108/SBM-05-2014-0027}

\bibitem{eco}
\textsc{Leeds, M. and Von Allmen, P} \textit{(2018).
The Business of Sports; ISBN-13: 978-11380521617}


\bibitem{david}
\textsc{David J. Berri and Martin B. Schmidt} \textit{(2006).
On the Road With the National Basketball Association's Superstar Externality; Journal of Sports Economics; 10.1177/152700250527509}


\bibitem{corr}
\textsc{Cohen, Jacob, and Cohen, Patricia} \textit{(1975).
	Applied multiple regression/correlation analysis for the behavioral sciences; Lawrence Erlbaum Associates ; distributed by Halsted Press Division of John Wiley Hillsdale, N.J. : New York}




\end{thebibliography}
\end{document}